# A classification
# of natural and social distributions
# Part one: the descriptions

L. Benguigui and M.Marinov
Israel Institute of Technology
Solid State Institute
32000 Haifa
Israel

> *Some time ago it occurred to the author that we might learn much about various social, economic, and political problems if, instead of viewing man as "God's noblest creation", we studied human group-behavior with the same ruthless objectivity with which a biologist might study the organized activity of ant hill, or a bee-hive or of a colony of termites.*
> *(G.K.Zipf in "National Unity and Disunity" p.1)*

ABSTRACT

This paper presents an extensive survey of regular distributions in natural and social sciences. The survey includes studies from a wide scope of academic disciplines, in order to create an inventory of the different mathematical functions used to describe the distributions. The goal of the study was to determine, whether a unique function can be used to describe all the distributions (universality) or a particular function is best suited to describe the distributions in each specific field of research (domain universality). We propose a classification of distributions into eighth different categories, based on three representations: the Zipf representation, the cumulative density function (CDF) and the probability density function (PDF). In the 89 cases included in the survey, neither universality nor domain universality was found. In particular, based on the results of the survey, the claim that "power law provides a good description for majority of distributions" may be rejected. Only one third of the distributions in our survey are associated with power laws, while another third is well described by lognormal and similar functions (Dagum, Weibull, loglogistic and Gamma functions). We suggest that correct characterization of a distribution relies on two conditions. First, it is important to include the full range of the available data to avoid distortion due to arbitrary cut off values. Second, it is advisable to display the data in all three representations: the Zipf representation, the CDF and the PDF.

INTRODUCTION

There is a multitude of phenomena in very different disciplines, for which the size distribution is known to display an apparent regularity. Some of the examples include populations of cities, frequencies of words in a text, incomes of individuals and firms in a country and number of people speaking the same language. In nearly every academic field, the scholars have been searching for a statistical formalism to describe the observed regularity. It is assumed, however, that not all distributions are regular enough to be studied and in some cases the regularity may be somewhat contrived.
Some of the early observations were made a relatively long time ago (see Par (1985) and Batty and Shiode (2003) for these precursors), but it was Zipf (1932, 1935, 1940, 1949), who has popularized the notion of "regularity in the distribution of sizes". Zipf is well known for his studies of the size distribution of cities, incomes and word frequencies. In these three fields, he proposed the graphical representation of distribution, now often called "the Zipf representation", in which one plots the logarithm of sizes (i.e populations of cities, incomes or frequencies of occurrence of words) versus rank. According to Zipf, in many cases the result is approximately a straight line, rendering these distributions to be power laws. The case of cities is particularly appealing, since the exponent found by Zipf is exactly -1 and thus the regularity is called "Zipf's law". In his writings, the author himself used the expression "harmonic series", rather than "power law". Effectively, the normalized sum of the items is: 1+1/2+1/3+1/4…, i.e. the harmonic series. At the end of his book *National Unity and Disunity* he writes: *"Why should nature be so infatuated with this harmonic series (if nature is)...the author hopes to deliver this theoretical proof of the inevitability of his series"*. In fact, Zipf believed, despite not being able to prove it, that many other phenomena could be described by his "series". Today one can rephrase this belief as follows: "Almost all distributions in social and natural sciences are regular and may be described rigorously by a mathematical expression".
Since Zipf, much effort was devoted to a description of regular distribution by a power law with a particular exponent. When the exponent is different from -1, the regularity is generally referred to as "the Pareto law" (Kleiber and Kotz, 2003).
In this article, we shall use the terms "power law" for all the cases (i.e. for Zipf or Pareto laws).
It is remarkable, that the impression one gets from numerous papers (Newman (2005), Pinto et al (2012)) is that in most fields of research, most frequent and most successful description of the size distributions is the power law. However, recently Perline (2005) challenged this opinion and proposed to distinguish between several cases. According to Perline, the first case includes the "Strong Power Law", when the complete distribution is a power law. The second case is the "Weak Power Law", when only part of the distribution can be convincingly fitted by a power law. Finally, "False Power Law" is the case, when a "highly truncated part" of the distribution may be approximated or mimicked by a power law. The typical example of this third case is the lognormal distribution, for which the upper tail can be approximated by a power law, although it is not rigorously a power law. Finally, Perline concludes that genuine power law cases are rather rare. A similar sentiment is expressed by Li (2005) in his paper *Zipf's law everywhere,* suggesting that care must be taken, while deciding whether a distribution is Zipfian or not.

In this paper, we undertake a broad survey of studies, in order to create a catalog of different functions proposed by researchers in various fields of natural and social sciences to describe regular distributions. The purpose of the survey is to determine, which functions are most commonly used, whether in some fields a unique form is successfully applied and finally - whether the power law is really the most appropriate function, as claimed by many authors.

The literature on this subject is truly huge. The studies are published in a large number of journals as the researchers belong to different disciplines, although they study the same phenomenon (for example, in the case of cities, geographers, economists, physicists and statisticians have all contributed to the debate). The survey presented here is by no means exhaustive, but in our opinion, it is large enough to give a detailed picture of the phenomenon. We would like to apologize to the authors, whose work is omitted from the survey. Below, we discuss the criteria used to select the studies to be quoted in the survey, such as, for example, a graphic representation of the data.

Much of the academic debate focuses on the choice of mathematical expression, which can best describe the size distribution of entities. Generally, the researchers can choose between three different representations of a distribution, each with its advantages and limitations. In the next section of the paper, we briefly discuss the three representations. We then propose a general classification of distributions into eight types, based on the different types of representations. We suggest that this classification will allow us to distinguish the cases of true power from the "weak" or "false" power laws, which may not be power laws at all. Subsequently, we discuss the application of the classification system to the survey data presented in the Appendix.

Finally, we would like to mention, that this paper does not discuss the different models, which were proposed in order to explain the distributions. We intend to do that in another forthcoming paper.

## THE THREE REPRESENTATIONS

There is nothing new in this section of the paper; nevertheless, it is useful to recall the three representations. First, we mention that in almost all the studies, it is assumed that a continuous function can be used to describe the distributions, despite the fact that in many cases, they are discrete distributions by definition.

1. The (S,R,) representation or the Zipf representation

We begin with the simplest representation, based on sorting the sizes S from the largest to the smallest. In sorting, each size is assigned a rank R, the largest receiving the rank R =1, the second - the rank R = 2, the third - the rank R = 3 and so on. It is difficult to plot S as a function of R because the values of R and S may be very large, so in it preferable to plot Ln S as a function of Ln R. At this point, a decision must be made: are the points in the graph aligned enough to be fitted by a mathematical expression? If the answer is yes, one speaks of the (S,R) representation.

2. The cumulative representation

Consider a value $R_1$ of the ranks corresponding to the size $S_1$. The number $R_1$ is the number of items with sizes equal or larger than $S_1$. Normalizing the ranks R to 100, $P_1$ (the normalized value of $R_1$) is the percentage of items with size larger than $S_1$. It results that the function P(S) (normalized ranks as a function of size) is the decreasing cumulative function (the CDF function). There is also the increasing cumulative function $C(S) = 100 - P(S)$.

3. The density representation

Suppose that one wants to know, how many items are between two values of the sizes or between S and S+ΔS? Supposing that the distribution is dense enough, so that the points in the P(S) curve are close enough to be well approximated by a continuous line, the answer is the quantity |dP/dS|ΔS or (dC/dS)ΔS. The function D = |dP/dS| is the probability density function or the PDF. If the distribution is not dense enough, the PDF may be represented by a histogram.

Usually, in a single study, only one of the three representations is used. As we will demonstrate below, to have a complete characterization of a distribution, inspection of all three representations is necessary.

THE CLASSIFICAIONS OF THE DISTRIBUTIONS

The following classification is based on a simple observation: the graph Ln S versus Ln R (S,R representation) for a very large number of distributions may have four different shapes: 1) a parabola-like shape with axis parallel to the x axis; 2) a parabola-like shape with axis parallel to the y axis; 3) a straight line; 4) a curve with an upward curvature.

It was already proposed (Benguigui and Blumenfeld-Lieberthal (2007)), that it is possible to approximate the first case by the following expression (y = ln S, x = ln R):

$$y = d + m(b - x)^{\alpha} \qquad (1)$$

where d, m and b are three parameters and α a positive exponent smaller than 1.

The second case can be approximated by the following expression:

$$y = d - m(b + x)^\alpha \qquad (2)$$

where, as above d, m and b are parameters and α an exponent larger than 1.

The third case of a straight line may be deduced from (1) or (2) by defining α = 1. This corresponds to a power law and m is the exponent of the power law.

The last case can be described by (1) with α < 0.

Table1 shows the three representations of the eight types of distributions.

The case, where α < 1, corresponds to three distributions: A1, A2 and A3. Which of the three – it will depend on a certain condition relating the parameters d, m and b.

The A1 distribution is characterized by the infinite slope at the maximum value of R, in the curve (Ln S versus Ln R), the CDF curve has a null slope for the smallest value of S and consequently the PDF curve exhibits a maximum.

The A1 type corresponds to several functions with a maximum, among them: the lognormal, the Dagum function (with values of the parameters giving the maximum), the Gamma distribution, the Weibull distribution and the log-logistic distribution. Some of these functions can be approximated by a power law in the upper tail, but in the definition of Perline (2005), these are "False Power Laws".

The A2 type has a very large value of the slope (but not infinite) in the curve (Ln S versus Ln R) for the maximum value of R, while the A3 has a small slope at the same point. The two types are very different. For the A2 type, the CDF and the PDF are finite for the small sizes, but for the A3 type, the CDF is finite and the PDF has a pseudo-divergence at small sizes. For these two types, the upper tail cannot be approximated by a power law. An example of the A2 type is the exponential distribution and that of the A3 type is the stretched exponential.

The distribution B is a straight line i.e. a power law (α =1) and m is the exponent of the (S,R) representation. If 1/m is the exponent of the (S,R) representation, the exponent of the CDF is m and that of the PDF is 1 + m.

The distributions C1 and C2 correspond to the case, where α > 1. Differentiation between them is related to a complicated condition for parameters d, b and m. Remarkably, the PDF of the C1 type is a straight line, making it impossible to differentiate from the power law. However, the Ln S versus Ln R is not a straight line, as in the case of the B type. The Zipf-Mandelbrot law is an example. The C2 type is characterized by the approximate behavior as a power law in the upper tail. One example of this type is the Simon model.

Type D corresponds to α < 0, only the small sizes can be approximated by a power law. Finally, type E represents a situation, when the distribution cannot be described by a single mathematical expression, but by a superposition of functions. For example, such distribution is manifested by a combination of two linear graphs, representing two power laws for the upper and the lower ranges of the distribution.

**Table 1: Types of distributions in three representations**

| Type | Rank Size | Cumulative | Density Function |
|---|---|---|---|
| **A1** | 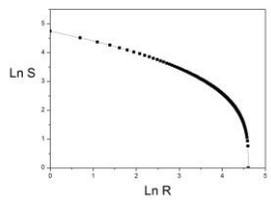 | 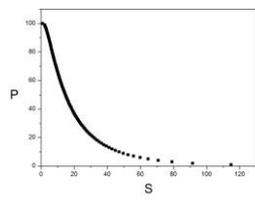 | 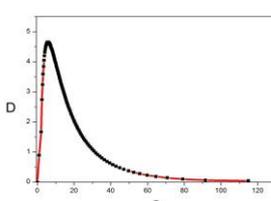 |
| **A2** | 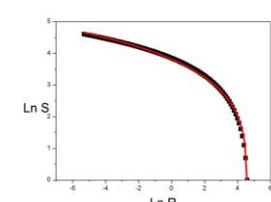 | 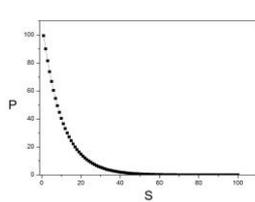 | 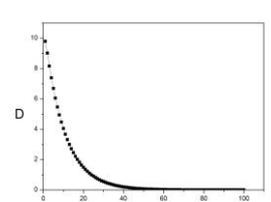 |
| **A3** | 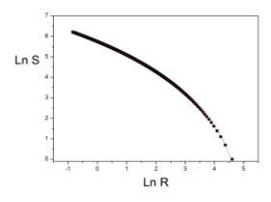 | 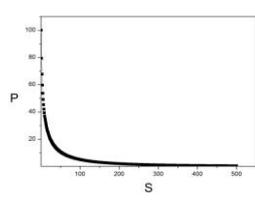 | 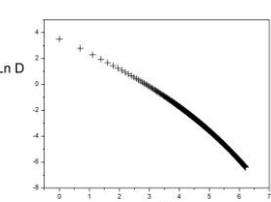 |
| **B** | 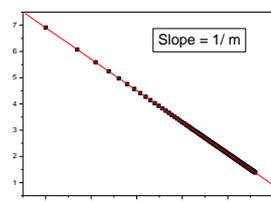 Slope = 1 / m | 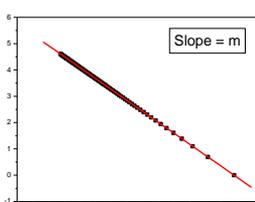 Slope = m | 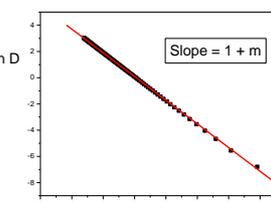 Slope = 1 + m |
| **C1** | 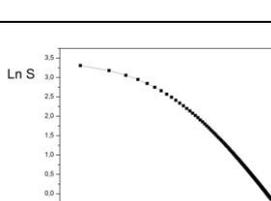 | 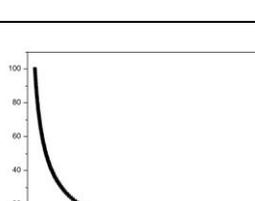 | 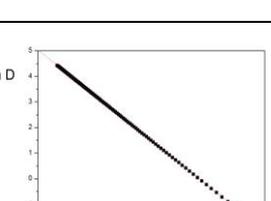 |

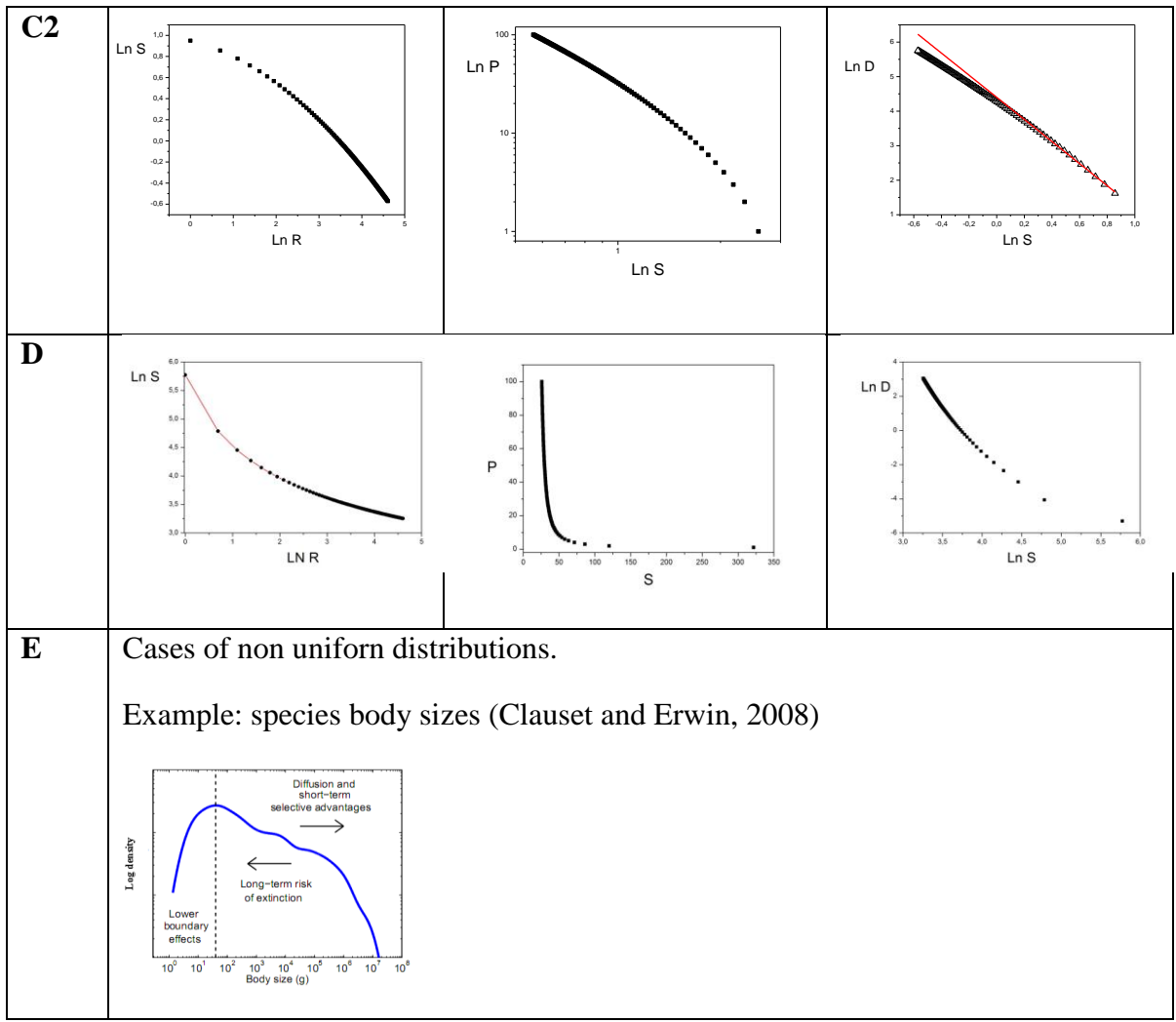

One can make the following observations concerning the classification:
1. Generally, we propose to consider a distribution to be a power law, if both the (S,R) and the PDF are well described by a power law (a linear graph in the log-log plot). To reach this conclusion it is not enough to look at the (S,R) representation – as was done by many scholars, eager to find "the Zipf law", but at the PDF as well. For type B the three representations are all linear, making it a clear-cut case of power law. On the other hand, for C1 the PDF is the same as for power law, but this is not so for the (S.R) and the CDF representations.
2. The distinction between the types of distributions cannot be made based on the (S.R) representation alone. The types A1 and A2 have a very similar graph of Ln S versus Ln R; nevertheless, they are very different distributions as mentioned above. Same goes for the types C1 and C2.

## THE RESULTS OF THE SURVEY

The survey includes a wide variety of studies in different spheres of knowledge. Firstly, we extracted the data from preceding surveys: Laherere and Sornette (1998), Newman (2005), Clauset et al (2009), Maruvka et al (2011), Martinez-Mekler et al (2009) and some others. Secondly, we carried out an extensive bibliographic research, the results of which are summarized in the Appendix. In the table in the Appendix, we display the curves from the original papers, the type of representation chosen by the authors, as well as the type of distribution according to our proposed classification. We consider a distribution to be type B only for the "Strong Power law" cases, i.e. when the complete distribution is well approximated by a linear curve. The survey only includes studies, which display a graph for at least one of the three representations. We do not discuss the different statistical methods used for the determination of the functions, although this too is not a trivial process (Clauset et al, 2009).

The survey includes 89 studies, which are divided into a number of broad categories. Table 2 shows the total number of cases for the different types of distributions. Apparently, types A1 and B are the most common, with a very similar number of cases. Together they comprise more than half of all the cases in the survey. The other cases (A2, A3, C1 + C2) are much rarer, with approximately the same probability of appearance for C and A3. The cases of C1 and C2 are united into one category, because the distinction between them is not always straightforward: the (S,R) representation alone is not sufficient to determine, whether a distribution is C1 or C2. Finally, the D cases are very rare: we found only two examples. In some cases, in order to verify, that the distribution cannot be fitted by a uniform mathematical expression (type E), one needs to apply elaborate methods of analysis.

The functions used for a description of the A1 type are as follows: lognormal, Dagum, Weibull, log-logistic, double Pareto lognormal (Reed (2002)) and others (Ramos et (2013)). All of the above are characterized by a PDF with a maximum and an upper tail approximated by a power law. There are only few cases, in which the authors undertook a comparison between different functions, suitable for description of the distribution. In most cases, such a comparison would require a complicated procedure, since the differences between the options are small and only a good model could distinguish between them (Quandt (1964)).

**Table 2: Statistics of distributions in literature**

| Percentage of cases | Number of cases | Type |
|---|---|---|
| 31.5% | 28 | **A1** |
| 4.5% | 4 | **A2** |
| 15.7% | 14 | **A3** |
| 29.2% | 26 | **B** |
| 13.5% | 12 | **C** |
| 2.2% | 2 | **D** |
| 3.4% | 3 | **E** |
| 100% | 89 | **Total** |

Three mathematical expressions correspond to the A2 type: exponential, q-exponential and the Dagum function. The best representation for this type is either the CDF or the PDF. Unfortunately many researchers prefer to use the (S,R) representation and their results are not conclusive.

The A3 type is represented by the stretched exponential and by the Kumer function (Maruvka (2011)).

Apparently only the Zipf-Mandelbrot distribution represents the case of type C1. We recall that the original Zipf-Mandelbrot function concerns the (S,R) representation and not the CDF as some researchers indicate (Mandelbrot (1953)).

The type C2 is represented by several functions: the Simon model and some cases of the Kummer function (Maruvka (2011)).

The first important conclusion stemming from our survey is that there is no discipline or sphere of research, in which all the distributions may be described adequately by a uniform function. The second conclusion is that, although power laws appear in almost all the spheres of research and despite their popularity, they are not the most common type of distribution.

An additional remark is due concerning the lognormal distribution. It was pointed out that the behavior of the lognormal for the large values is power law-like (although it is not rigorously true even asymptotically). Moreover, the exponent for this upper tail of the lognormal is equal -1, the Zipf exponent. In many cases, the power law behavior was identified essentially for larger items, while ignoring the bottom part of the distribution, the Zipf exponent found only above an arbitrary cut off value. Ekhout (2004) and others (Batty and Shiode (2003)) observed that the small items are accurately described by a lognormal function. It was also suggested (Matthews et al.(2002)) that it is difficult to discriminate between the functions giving a A1 type behavior (i.e. the PDF exhibits a maximum) such as Lognormal, Dagum, Weibull and log-logistic functions. We suggest that, at least in some cases, the ubiquity of the exponent -1 is only the ubiquity of the lognormal distribution. Furthermore, it is possible that all the cases of A1 can be classified as examples of lognormal distribution (as well as some of the cases of the Weak Power Law), thus supporting the claim that the lognormal distribution is the most common and widespread distribution.

In the next sections, we perform a more detailed analysis of our results for a number of disciplines, for which there is a large amount of data: cities, linguistics, earthquakes and economics.

**Cities**

According to Zipf, there is a Law that the sizes of cities of a given country can be ordered following their ranks, such that $S = A/R^q$, where q is approximately (or exactly) equal to 1. It has often been stated (implicitly or explicitly), that this law concerns only the "upper tail" of the distribution. In other words, a cut off S' is introduced, so that only cities of size larger than S' are included in the distribution. The choice of this cut off was for a long time a matter of subjective preference (see

Par (1985), Guerin-Pace (1995), Rosen and Resnick (1980)) for the discussion of the cut off size, that will produce the power law distribution). In fact, researchers were faced with two interconnected questions: "what is the minimum size of a city" (the cut off size) and "what is a city". Evidently, the values of q depend on the answers to these questions (Rosen and Resnick (1980), Nota and Song (2008), Scaffar (2009)). In consequence, the interpretation of the exponent q was the object of intense discussions. Finally, the following consensus was reached: the "upper tail" of the city size distribution is either an exact power law or a very good approximation, with q between 0.85 and 1.15 (Par (1985), Berry (2011), Pumain (1991), Batty and Shiode (2003), Gabaix and Ioannides (2010), Rosen and Resnick (1980)). Subsequently, researchers attempted to correlate the value of the exponent q and its variation with time to economic and geographic processes.

At the same time, other suggestions were made, which cast doubt on the quality of this approximation (Alperovich (1984)) or even on the concept of fitting this type of distribution with a power law function (Par and Suzuki (1973), Eechkhout (2004), Perline (2005)). It was claimed that, taking into account the complete distribution, the adequate function should be the lognormal and consequently the upper tail is that of the lognormal function, approximately a power law with exponent 1. It was shown that a superposition of several lognormal functions gives an approximate power law (Par and Suzuki (1973), Perline (1996)). Laherere and Sornette (1998) proposed the stretched exponential as a good fit. Benguigui and Blumenfeld-Lieberthal (2007) used the above classification to show, that there are countries with the exponent $\alpha > 1$ (C1 or C2 type), others with $\alpha = 1$ (power law or B type) and one country (China) with $\alpha < 1$ (A1 type)).

Reed (2002) proposed the model called the double Pareto lognormal (DPLN), which is qualitatively similar for cities. The interesting point is that for the small cities, the distribution is an increasing power law but for the larger ones it is a decreasing power law (hence the name of double Pareto). These functions are (same as the lognormal) of the type A1 with a maximum in the PDF.

It is likely, however, that the other types may adequately describe city size distributions in some countries. For example, in the case of India, the (S,R) graph is a compound of two decreasing straight lines, showing that the small cities cannot be described by the A1 type function (Benguigui and Blumenfeld-Lieberthal (2011)).

Eeckhout (2004) studied the USA Census 2000 using the full range data and claimed that the complete distribution is lognormal. A power law may approximate the upper tail, but the best function for the full distribution is the lognormal. These results were contested by Levy (2009) and by Malevergne et al (2011): the small cities are lognormal, but the upper tail is a power law. This is a late confirmation of the suggestion made by Par and Suzuki in 1973. The work of Malavergne et al (2011) defines the cut off city size, which will produce the exact power law result. This problem was studied in great detail by Clauset et al (2009) in order to define exactly what part of a distribution is a power law.

Gonzalez-Val et al (2013) studied the cases of France, Spain and some other countries belonging to OECD. After comparison between several functions (lognormal, DPLN and log-logistic) they concluded that the best fit was obtained with the DPLN, as proposed by Reed (2002).

Finally, one can quote the work of Ramos et al (2013), who examined the city size distribution of the USA. Their suggestion was to distinguish between three groups of cities: the small cities are described by an increasing power law and the large cities - by a decreasing power law. The intermediate size is given by the Singh-Maddala (1976) function. Conequently, this decription necessitates two cut offs, one for each of the two power laws.

In conclusion, it is well established that the upper tail of city size distribution is either a Strong Power Law or a good approximation. However, these two possibilities are not equivalent: the first one will be classified as the type E (since only the upper part is a power law when for small sizes an another function is good); while in the second, the complete distribution is described by one function which is a power law only asymptotically (even if the approximation is very good). The lognormal and DPLN are examples of such a case. There are also some exceptions to this conclusion, when few of the largest cities cannot be included in the distribution.

The description for the distribution of small cities is less clear. There is evidence for some countries that the PDF tends to zero for the smallest sizes, and that this PDF has a maximum (A1 type), although, this is not necessarily a general phenomenon. It is possible, that in some countries the distribution is of the C1 or C2 type, with a pseudo divergence for the small cities. Many more studies were devoted to the city size disribution in the US, than in other countries, thus further research is required to gain a general understanding of the phenomenon.

**Earthquakes**

Seismology research was one of the first to recognize regularity in the size distribution of its parameters. The relationship between the magnitude and the total number of earthquakes was described by a power law as early as 1944 by what is today well known as the Gutenberg-Richter law (Gutenberg and Richter, 1944). The law states that:

$$Log(N) = a - bM,$$

where N is the number of earthquakes of magnitude larger than M, "a" and "b" are constants. The constant "a" is associated with level of seismicity and "b" is generally close to 1 in seismic areas. The presence of power law in the distribution of magnitudes of earthquakes is typically associated with the characteristics of self-similarity and scale invariance. This connection between the power law in the distribution of magnitudes and the self-organized criticality phenomenon, is discussed by Turcotte (1999) and Turcotte and Malamud (2004). The authors use the cumulative distribution for their analyses, which indeed display a good agreement with the Gutenberg-Richter law. Similar results are obtained for distributions of landslides, which can be regarded as an event caused by earthquakes, and wildfires - both of these displaying self-similar behavior.

Much of the debate in the seismological literature has revolved around the value of "b" and it has been suggested, that its departure from the constant value signifies that there are in fact fewer small earthquakes, than predicted by the power law. In general, not many studies challenge the claim that the power law holds for seismic data, that being the basic assumption. This is surprising, in view of the fact that in many cases upon quick inspection it becomes clear, that the empirical data provided by the authors agrees with the power law only partially.

For example, Abercrombie (1996), who analyses the data for earthquakes in Sothern California, after a brief discussion concludes that the values of "b" are constant, however the cumulative distribution displayed by the author in figure 5 allows us to classify the distribution as type A1 or partial power law. Sornette et al (1996) discuss worldwide data set, which displays a transition between small and large earthquakes and with different values of "b" for each group (type E). The authors caution against the universal use of power law and suggest that the gamma distribution fits the whole range satisfactorily.

Lahererre and Sornette (1998) suggest that the stretched exponential can be used "as a complement to the often used power law" to describe the distribution of earthquake sized and fault displacements. However, the authors use only the Rank-Size representation, thus the comparison of the suitability of the different models is somewhat limited.

More recently, in his extensive review of power laws and Pareto distributions, Newman (2005) uses the earthquakes as an example of power law distribution. We suggest that the cumulative representation presented by the author is better described by type C1 distribution, rather than type B.

Finally, Abe and Suzuki (2005) studied the distribution of time intervals between seismic events in California and Japan and found that these too obey a power law. However, the cumulative distribution displayed by the authors can be classified as type A2.

**Linguistics**

The distribution of word frequencies in texts is primarily associated with the power law originally popularized by Zipf (1935, 1949). It is widely accepted that this law (or its extension proposed by Mandelbrot (1953, 1962)) generally holds for different languages and different texts, although it is recognized that in many cases it provides a good fit only for some sections of the distribution (Wyllys, 1981; Montemurro, 2001; Piantodosi, 2014). The scope of literature on the subject is enormous and the debate is ongoing, therefore we will only name a few examples of studies, which examine the form of word frequency distribution. Almost every study performed on the word frequencies since Zipf, seeks to rediscover the power law and attempts to find reasons for deviation from it, when such deviation is found. There were, of course, some attempts to find an alternative form. It has been suggested, for example, that the word frequency distribution may be described by log-normal distribution (Carroll, 1967, 1969) or by a generalized inverse Gauss-Poisson model (Sichel, 1975).

Montemurro (2001) emphasizes that in large text corpora, many words do not follow the same pattern and even for texts with high homogeneity, the less common words form a faster decaying tail, straying from the power law model. In linguistics, the use of "Rank versus Frequency" representation seems to be by far the most popular – perhaps, due to tradition established by Zipf himself. Based on this representation alone, we may classify the curves displayed in the paper by Montemurro as type C (C2).

In his recent overview, Piantodosi (2014) acknowledges, that "word frequencies are not actually so simple", but his conclusion is that the "large-scale structure" for many languages is still well described by Zipf's law. Despite the realization, that this method of analysis is somewhat limited, the author himself applies the Rank-Frequency representation almost exclusively, fitted by the Mandelbrot extension. We propose that these too can be classified as type C (C1).

In some cases, the very large size of text corpora or lack of homogeneity is the alleged reason for the deviation of the data from the expected power law pattern. For example, Dahui et al (2005) compares the frequency distribution of words in Chinese language for different historical periods and finds that only for some of them the data conforms to Zipf. The distribution, which does not follow the power law in this paper, is recognized as type D.

Sigurd et al (2004) tested word length and sentence length to verify an additional aspect of power law in linguistics. The authors recognize that the distribution conforms to power law only partially and propose a gamma distribution to as a better approximation. We identify the distribution of word lengths in the study as a type A1 distribution.

**Economics**

Majority of studies on size distributions of economical entities discuss the distribution of incomes. Zipf was one of the first to describe the distribution of incomes by a power law, similar to the law he discovered in linguistics. Later studies suggested that the power law holds for the higher incomes only, while various solutions were proposed for the remaining range of lower incomes. Montroll and Schlesinger (1983) and later Reed (2002, 2003), proposed the double Pareto function – 2 different power laws for the higher and the lower parts of the distribution (type E distributions). Some authors were satisfied with the "Weak Power Law" result (Okuyama et al, 1999), although it was evident that parts of the data depart from the strict power law pattern. Another approach suggests that a single function may be used to accurately describe the full range of incomes, but not a power law. Among these functions were the exponential (Dragulescu and Yakovlenko (2001)) – type A2, the stretched exponential (Laherere and Sornette (1998)) – type A3, the gamma distribution (McDonald (1984)) – type A1, the κ-generalized function (Clementi et al (2008)) – type A1 and the Dagum function (Dagum (1990), Lukasiewicz and Orlowski (2004))- type A1. Some other functions were formulated based on theoretical and computerized models (Yuqing (2007), Chatterjee et al (2004)). Researchers have also looked at distribution of wealth rather than incomes. Levy and Solomon (1997) found that the wealth of the richest people in the US is in good agreement with the power law – type B. Sinha (2006) discovered a weak power law in the distribution of wealth in India. In these cases, yet again, a truncated data set was used, including only "the rich tail" of the distribution.

The variability of functions, used by the researchers in order to describe the distribution of incomes, may lead to a conclusion that a universal function describing the full range of incomes in various countries simply does not exist. Each data set is unique and may have another mathematical form best suited to describe it. Here again we may raise the question, whether the expectation for the existence of universal rules (generally accepted in physics) is justified in the case of incomes and wealth distribution.

# CONCLUSION

We have presented a survey, which includes studies performed by researchers in a wide variety of disciplines, applying different approaches and methods and proposing sometimes very different theories in order to describe size distributions of natural and social phenomena. What unites all these studies is the idea, that there exists a regularity in these distributions, which sometimes is evident and at other times requires rigorous analysis, in order to be unveiled. In some cases, there is also a strong academic belief that there is an underlying force, which is essentially responsible for the uniform distribution, in other words – there is a reason for these regular patterns. Understanding the connection between the regular size distribution and the underlying forces creating them is the key to understanding the processes, which shape the reality surrounding us. On the other hand, the more skeptical or pragmatic approach would be to question the existence of any all-purpose functions, uniform regularities and all-embracing theories, which would provide simple answers to complex questions.

This survey is an effort to systematize the efforts in the many different fields by means of a relatively simple qualitative classification. Firstly, in is important to note that the evidence of regularities in the size distributions is rather overwhelming. The cases of type E are quite rare, so we might boldly proclaim that majority of the studied phenomena display a type of regularity in their distributions.

We divide the studied entities into two large groups: natural phenomena and human-related activity. We then determine which type of distribution it belongs to, based on its graphic representation. Based on the results of the survey, it can be stated that majority of cases belong either to type A1 or to type B distribution. The type B is the "Strong Power Law" and these are not the overwhelming majority, as was believed by many in the past, but also not the rare special cases, as was suggested by others. The other large group - type A1, includes a variety of functions, of which the most common and, perhaps, the most trivial is the lognormal. Despite the many efforts to characterize the distributions and to find the best fitting mathematical expression, there are not many studies that perform a thorough comparison between the different functions, suitable to describe the distribution at hand. Some such comparisons were carried out by Mitzenmacher (2004) and Lahererre and Sornette (1998), but many more are necessary in order to make more decisive conclusions. For instance, in the A1 category, there is a multitude of functions, proposed by the researchers and fitted by a variety of methods, but it would be interesting to test, whether all of these (or many of them) may as successfully be described by a lognormal function.

Another interesting question relates to the transition between the lognormal (or another A1 type function) and the power law. Under certain conditions a distribution, originally characterized as a lognormal, changes thus allowing the power law to emerge. Some such mechanisms have been proposed and some interesting models, resulting in transition between power law and other types of distributions, have been developed (Mitzenmacher (2004)).

In our next paper, we will attempt to create a general overview of the models, suggested by the scholars in the different fields in order to explain the regular distributions.

In addition, we would like to stress that the use of a single representation (especially the (S,R)) is usually not enough to identify the type of distribution. As we have shown, very different functions may appear very similar. Therefore, it is our suggestion, that analysis of the distribution be carried out in at least two different representations, allowing for the choice of the adequate function, best suited for its description.

Finally, we suggest that an open mind is kept regarding the expected regularity. We have seen how easy it is to fit an apparently non-linear curve with a straight line, or how convenient it would be to define a cut off value, which would render the remaining distribution a perfect case of Zipf's law. We must learn to be impartial and sometimes to trust the human eye more than the elaborate statistical methods. The proposed classification offers an easy visual key to recognition of the type of distribution, to be backed up (but not replaced) by a more sophisticated calculation.

.

# Appendix

In this appendix, we present 62 studies in which authors used various mathematical expressions to describe the size distribution of phenomena. Of all the studies included in this survey, 18 deal with natural phenomena and the rest – with some type of human activity. We would like to emphasize the extraordinary variety of disciplines, to which the studies belong. This includes distributions in almost every sphere of human activity (art, music, languages, violence, games, religion, internet and others). In the natural sciences, one finds examples in biology, earth sciences, chemistry and ecology. The survey spans 40 different phenomena, with some (such as cities, incomes and words in languages, for example) much more popular than others, therefore in some cases few studies of the same phenomena are listed.

The survey table includes the function chosen by the authors to describe the distribution, as well as its original graphic representation. Normally, the authors use only one of the three possible representations; two different representations of the same phenomena are a rarity. The type of distribution based on our new proposed classification is displayed in the right-hand column of the table. The distinction is made according to the graphic representation of the data in the original study. As we have discussed in the text, one representation is not always sufficient to make a definitive identification of the type of distribution. For example, the distinction between type B and type C1 is not possible on the basis of the (S,R) representation. Same thing applies to the types C1 and C2. However, the cases of ambiguity are not numerous and we believe that our conclusions remain valid: the types A1 and B are the most commonly observed, while other types of distributions are less common.

| Phenomenon | Authors/date | Suggested function | Representation | | Type of distribution |
|---|---|---|---|---|---|
| **Nature** | | | | | |
| **Earthquakes** 1. Seismic magnitude | Abercrombie, 1996 | Partial Power law | Cumulative 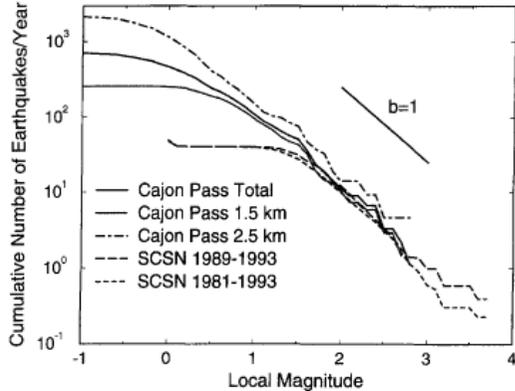 | | A1 |
| 1. Seismic magnitude | Turcotte and Malamud, 2004 | Power law | Cumulative 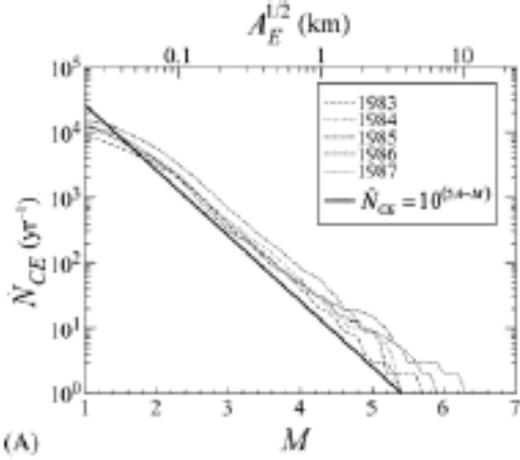 | | B |

| 1. Seismic magnitude | Lahererre and Sornette, 1998 | Stretched exponential | Rank size 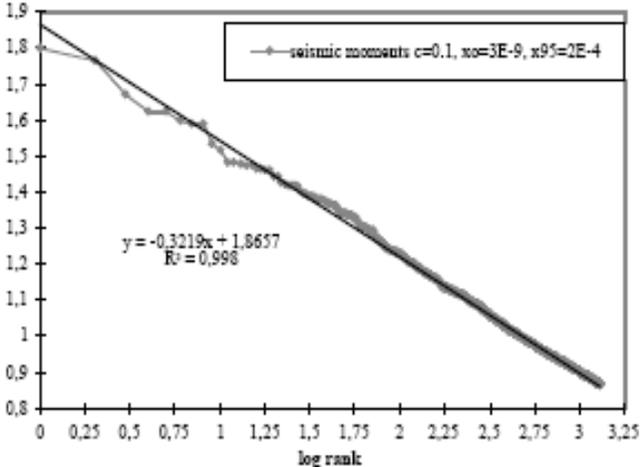 | A3 |
|---|---|---|---|---|
| 1. Seismic moment | Sornette et al, 1996 | Power law – 2 different laws: for large events and smaller events | Rank size 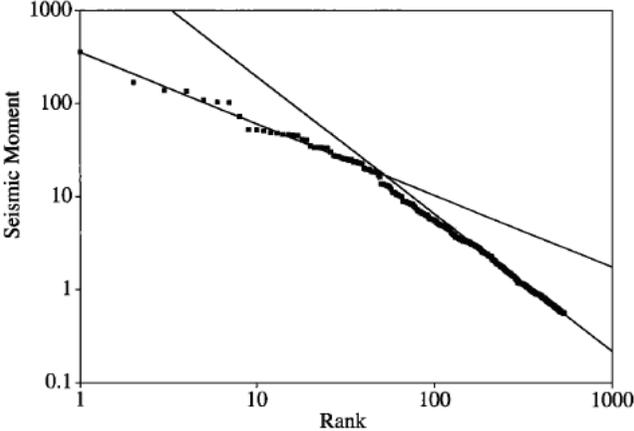 | E |

| 2. Time intervals between seismic events | Abe and Suzuki, 2005 | Zipf-Mandelbrot law, q-exponential function | Cumulative 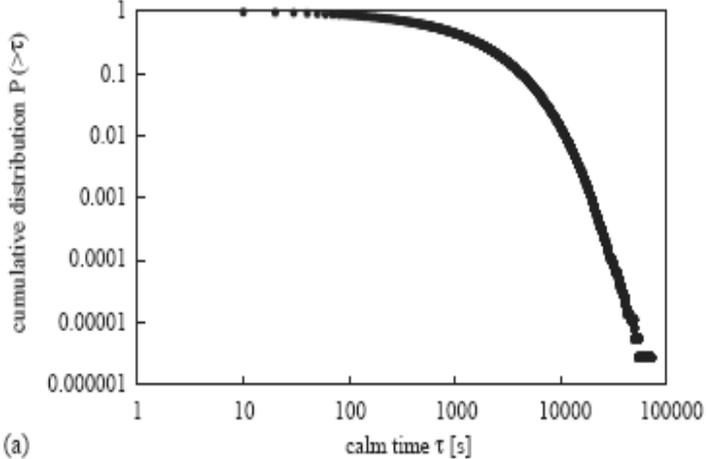 | A2 |
|---|---|---|---|---|
| **3. Landslides** | Guzzetti et al, 2002 | Partial Power law | Density function 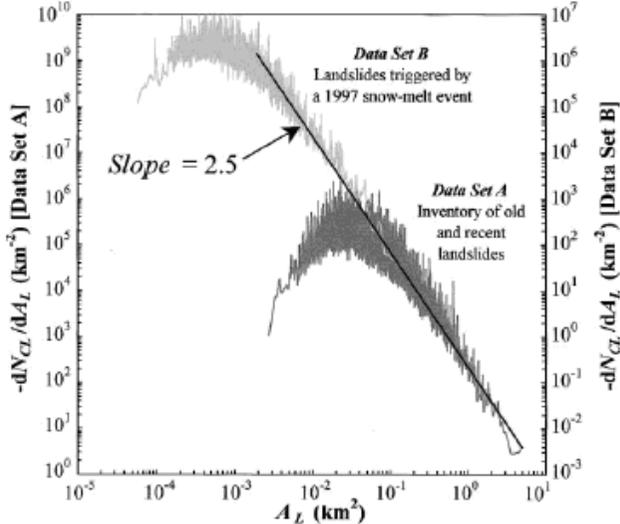 | A1 |

| 4. Wildfires | Malamud et al, 2005 | Power law | Cumulative 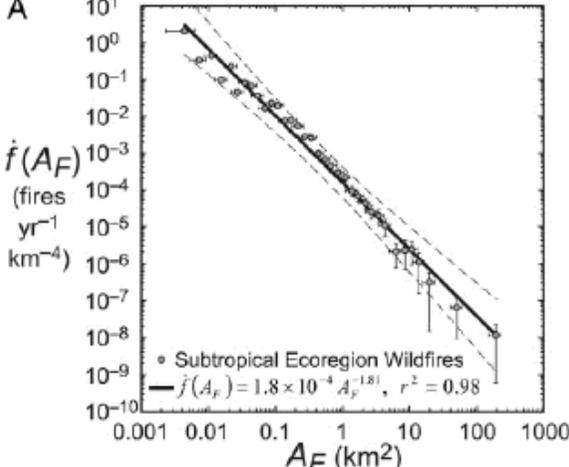 | B |
|---|---|---|---|---|
| 4. Wildfires | Turcotte and Malamud, 2004 | Power law | Density function 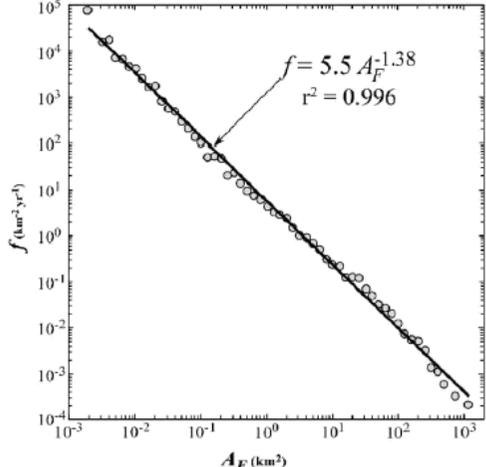 | B |

| 5. Lake size | Seekell and Pace, 2011 | Pareto for truncated data | Cumulative | B |
|---|---|---|---|---|
| | | | 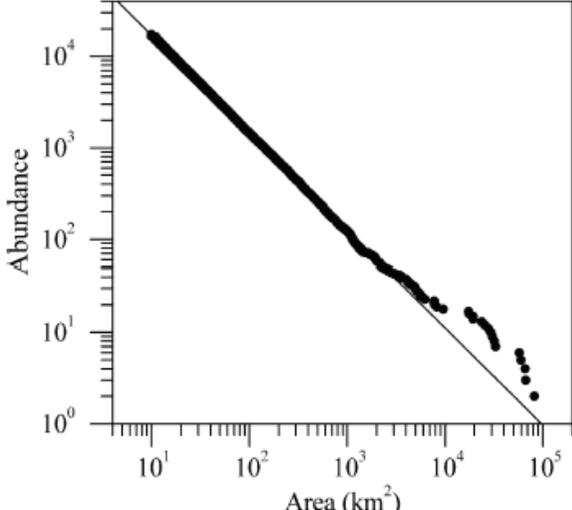 | |
| 6. Ozone level | Sexto et al, 2013 | Dagum $F(x) = \alpha + (1-\alpha)[1+(x/b)^{-a}]^{-p}$ | Cumulative 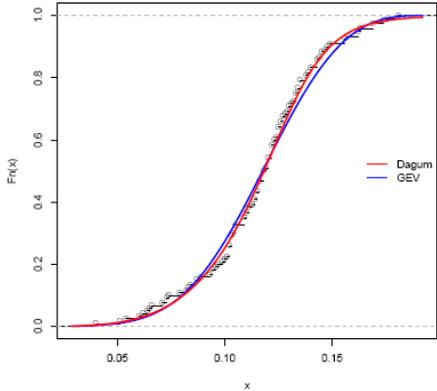 | A1 |
| **Biology** | | | | |
| **7. Epidemic events** | Rhodes and Anderson, 1996 | Power law | Cumulative 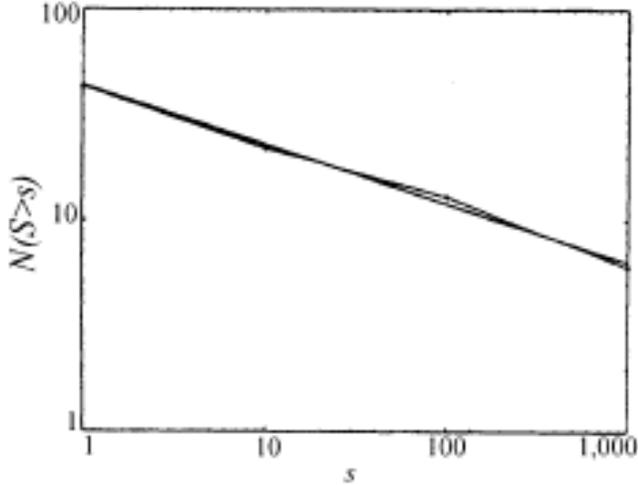 | B |
| **8. Water permeability in plants** | Baur, 1997 | Lognormal | Cumulative 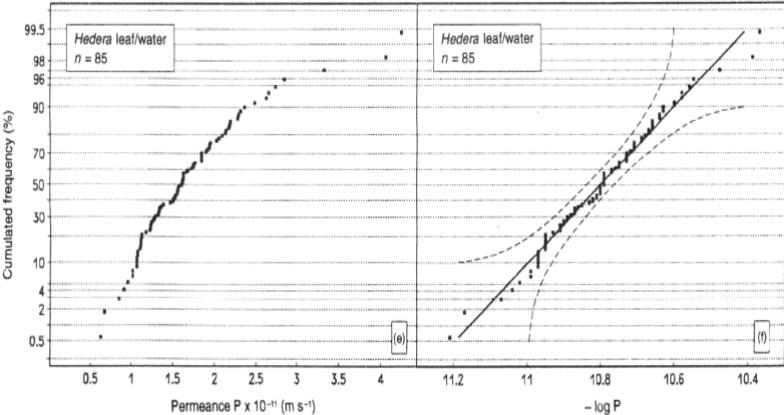 | A1 |

| 9. Onset of Alzheimer's disease | Horner, 1987 | Lognormal | Cumulative 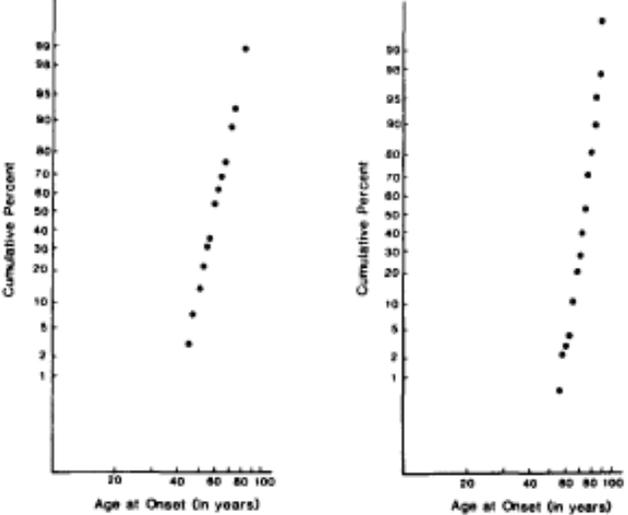 | A1 |
|---|---|---|---|---|
| 10. Species abundance | McGill et al, 2007 | Lognormal | Cumulative 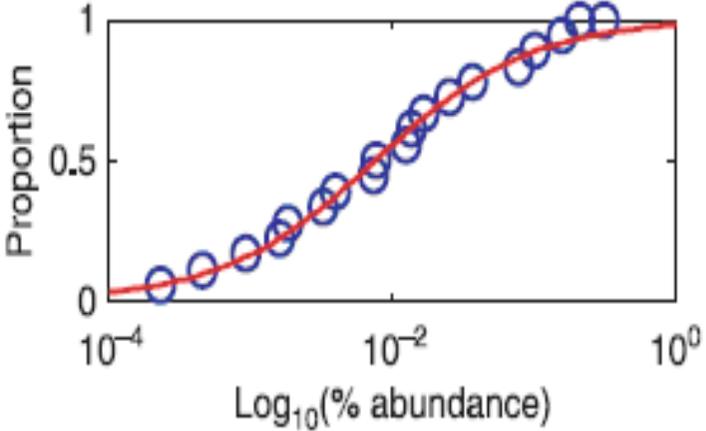 | A1 |

| 10. Species abundance | Maruvka, Kessler, Shnerb, 2011 | Kummer function | Density function 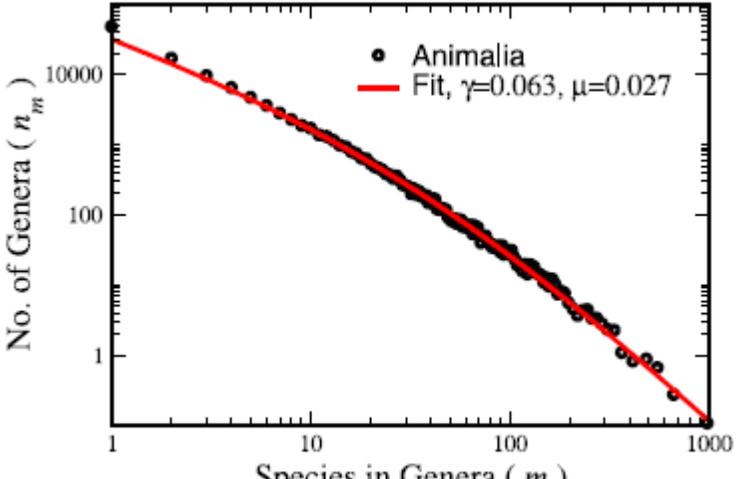 | A3 |
|---|---|---|---|---|
| **11. Area occupied by different species** | Martinez et al, 2009 | $f(r) = A(N+1-r)^b / r^a,$ | Rank –size 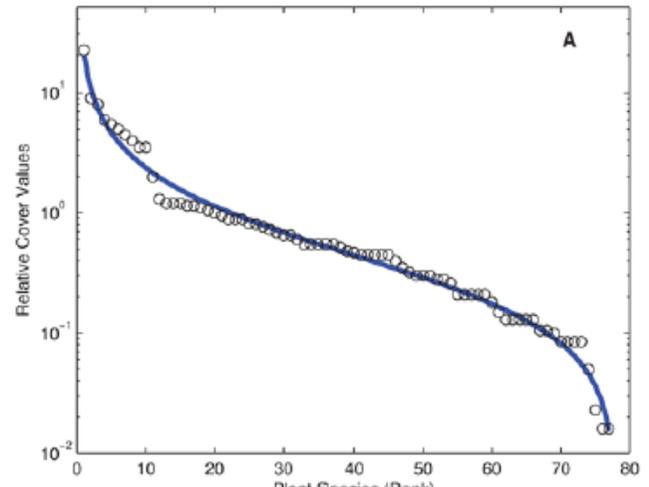 | A1 |

| 12. Body size of species | Clauset and Erwin, 2008 | A number of distributions | Density function 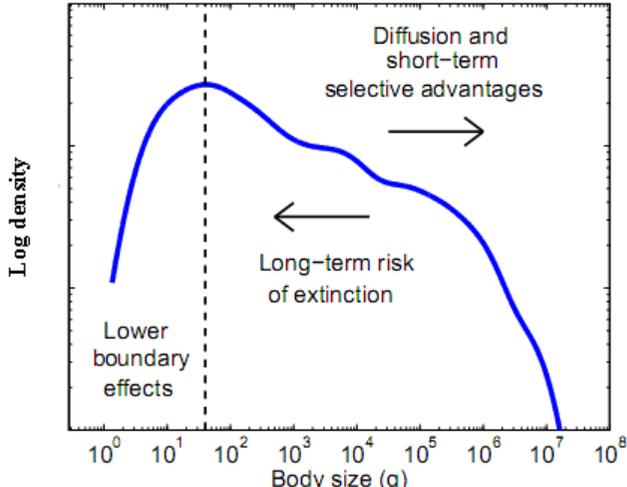 | E |
| --- | --- | --- | --- | --- |
| 13. Genetic sequences | Martinez-Mekler et al, 2009 | $f(r)=A(N+1-r)^b/r^a,$ | Rank-size 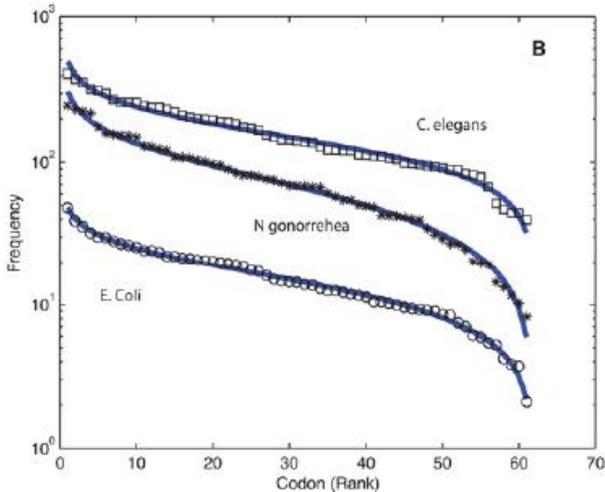 | A1 |

| **Human activity** | | | | |
|---|---|---|---|---|
| **Internet** <br> **14. Email addresses** | Ebel et al, 2002 | Partial Power law | Cumulative <br> 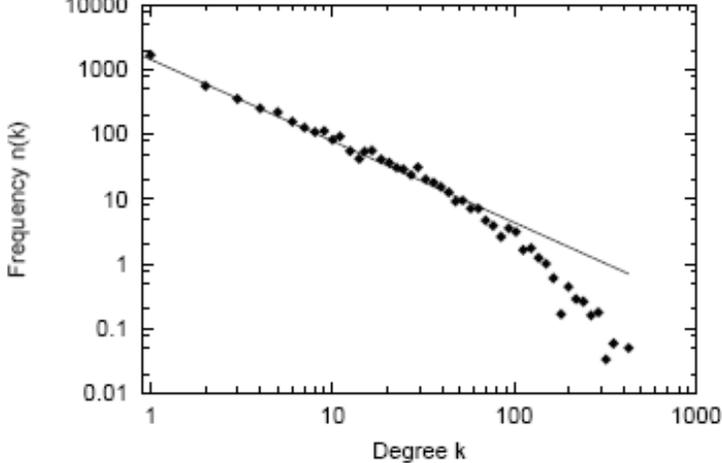 | A3 |
| **15. Links in sites** | Adamic and Huberman, 2000 | Power law | Cumulative <br> 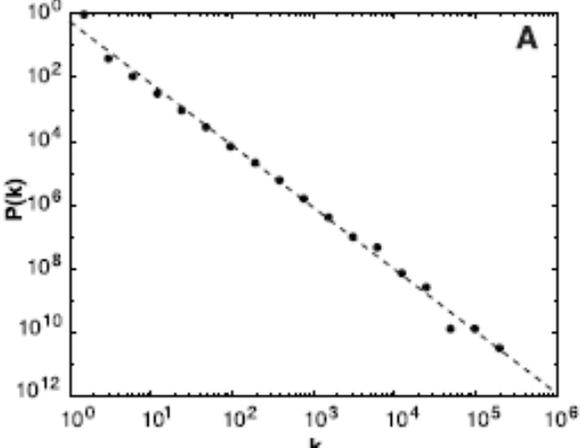 | B |

| 16. Download frequencies | Han et al, 2004 | Power law | Rank-size 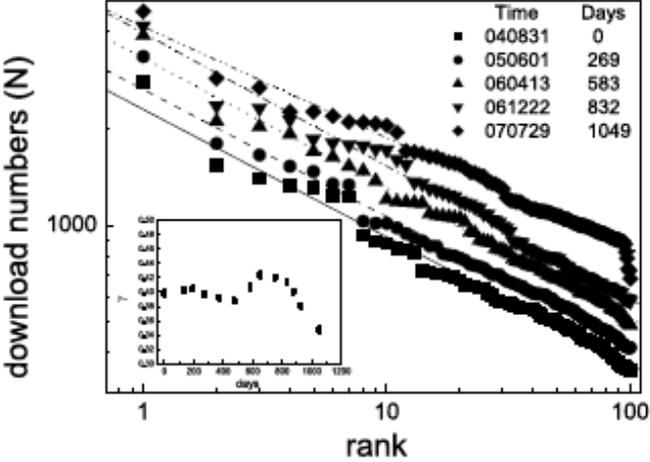 | B |
| --- | --- | --- | --- | --- |
| 17. Words on the www | Wei et al, 2005 | Partial Power law | Rank size 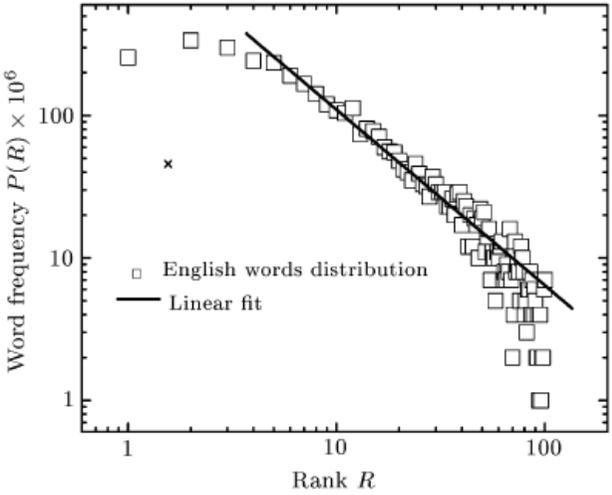 | A3 |

| | | | | | |
|---|---|---|---|---|---|
| **Violence** **18. Terrorist events** | Clauset et al, 2012 | Power law for Deaths Partial power law for Total | Cumulative 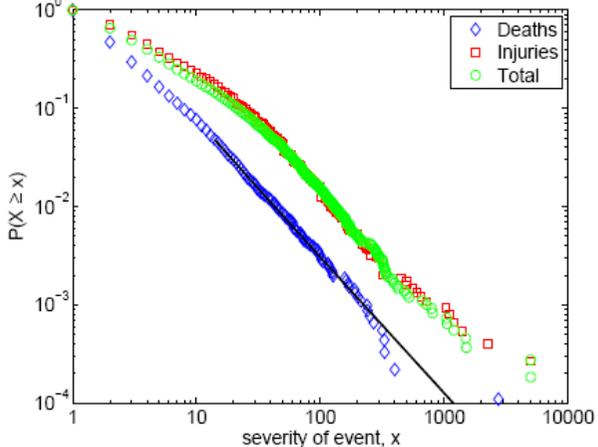 | | B A3 |
| **19. Inter-event time terrorist attacks** | Zhu et al, 2010 | Power law | Rank size 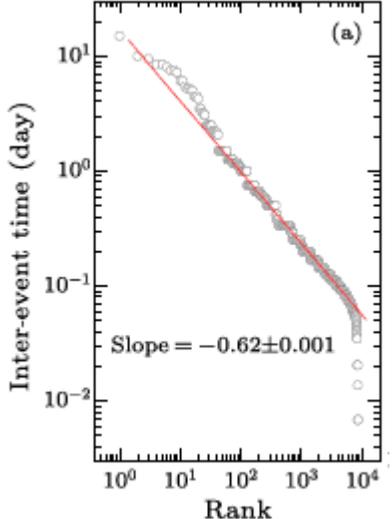 | | B |

| Other 20. Paper citations | Tsallis and Albuquerque, 2000 | $N(x) = N(1)\frac{[1+(q-1)\lambda]^{\frac{q}{q-1}}}{[1+(q-1)\lambda x]^{\frac{q}{q-1}}}$<br><br>Stretched exponential and power law only partial fit | Cumulative<br>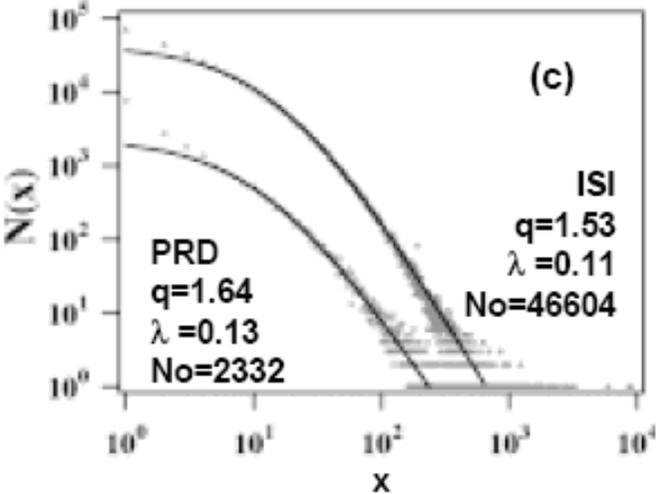 | A3 |
|---|---|---|---|---|
| 21. Religious adherents | Ausloos and Petroni, 2007 | Weibull or lognormal | Rank Size<br>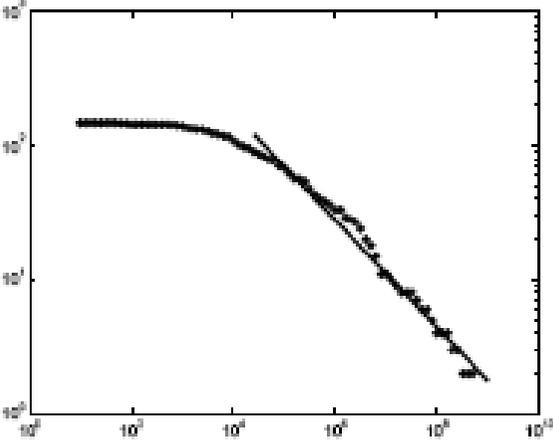 | A1 |

| 22. Family names | Miyazima et al, 2000 | Power law | Density function 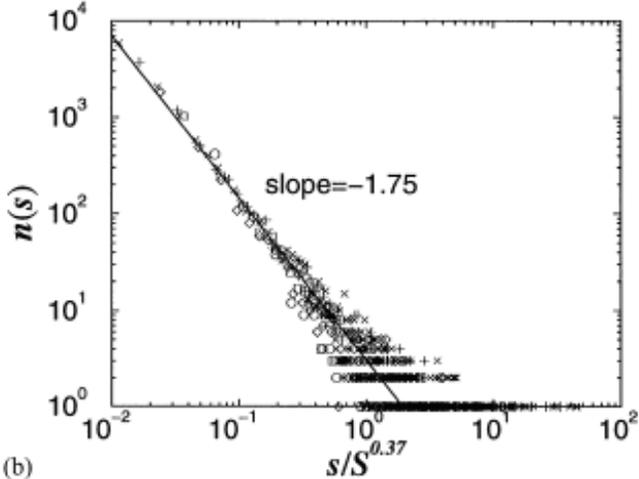 | B |
| --- | --- | --- | --- | --- |
| 23. Horse race bets | Ichinomiya, 2006 | Power law at X<100, Exp decay at X>100 | Cumulative 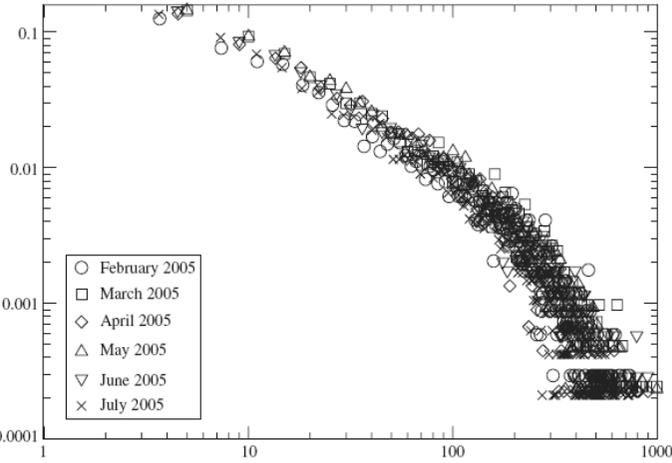 | A3 |

| 24. Chess openings | Blasius and Tonjes, 2009 | Power law | Density function 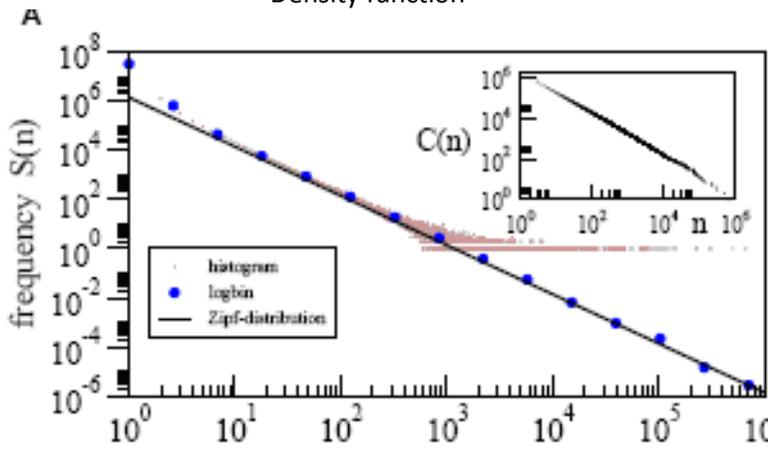 | B |
|---|---|---|---|---|
| 25. Art – size of geometrical shapes in art | Martinez-Mekler et al, 2009 | $f(r) = A(N+1-r)^b / r^a,$ | Rank size 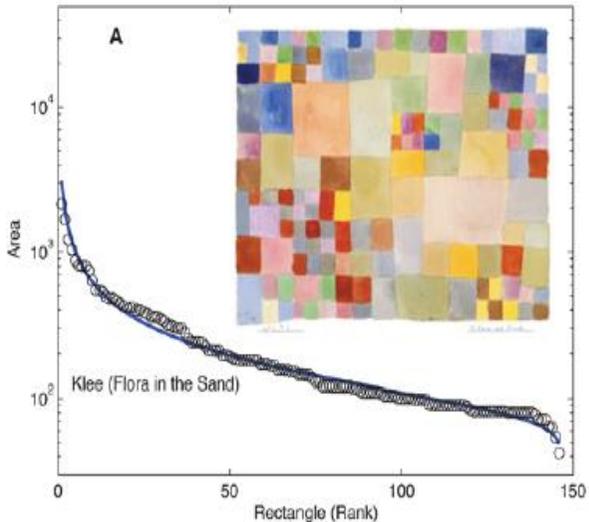 | A1 |

| 26. Music – Occurrence of notes in compositions | Martinez-Mekler et al, 2009 | $f(r) = A(N+1-r)^b / r^a,$ | 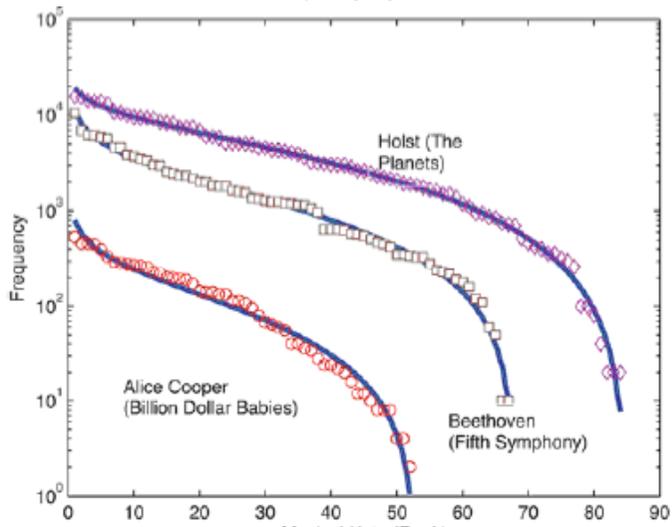 Rank Size | A1 |
|---|---|---|---|---|
| 26. Music | Martinakova et al, 2008 | | 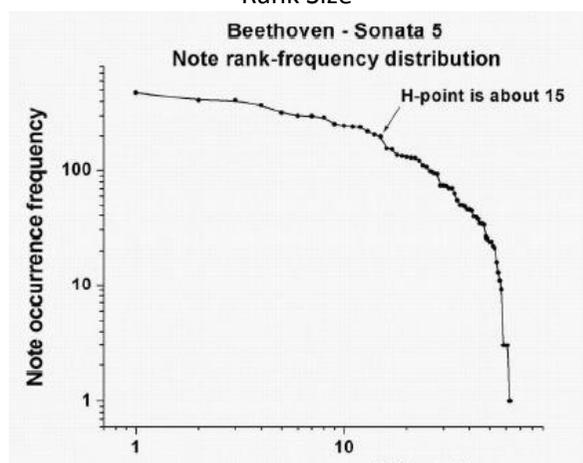 Rank Size | A1 or 2 |

| 26. Music | Zanette, 2004 | | Rank-size 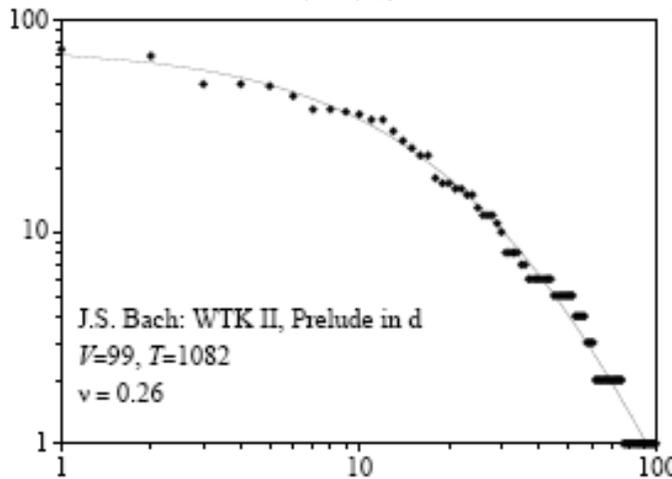 | C1 or C2 |
|---|---|---|---|---|
| 27. Football: number of goals by players | Malacarne and Mendes, 2000 | Power law | Density function 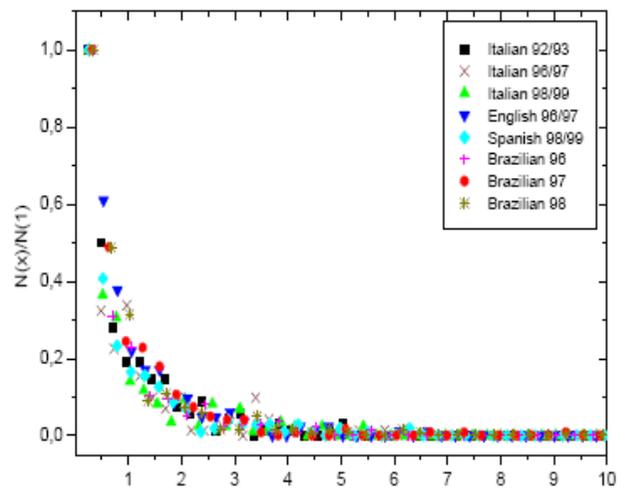 | B |

| 28. Movie actor collaborations | Martinez-Mekler et al, 2009 | $f(r)=A(N+1-r)^b/r^a,$ | Rank size 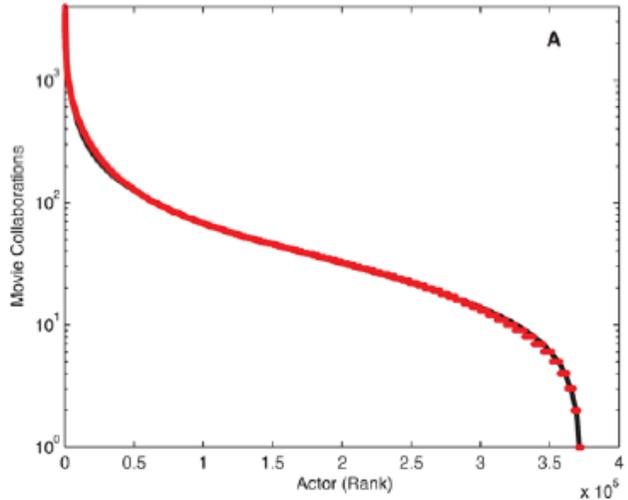 | | A1 |
|---|---|---|---|---|---|
| **Cities** | | | | | |
| **29. City Size distribution** | Eeckhout, 2004 | Lognormal, Zipf holds only in the upper tail | Density function 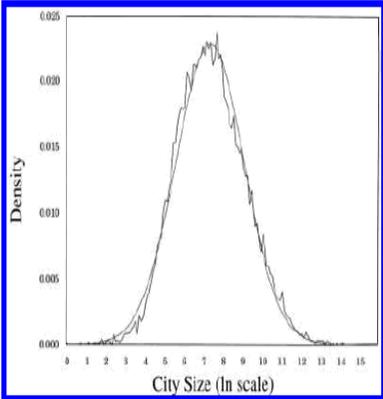 | Cumulative 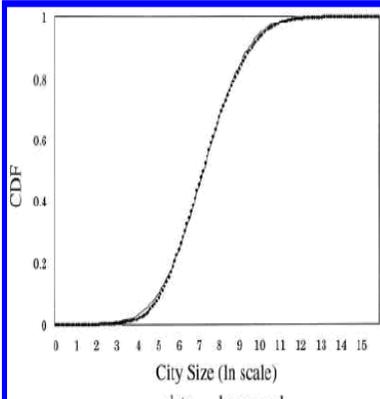 | A1 |

| 29. Cities | Benguigui, Blumenfeld-Liberthal, 2007 | $y = y_0 - a(b+x)^\alpha$<br>α>1<br><br>$y = y_0 + a(b-x)^\alpha$<br>α<1<br>y - ln(size), x – ln(rank), | Rank-size<br>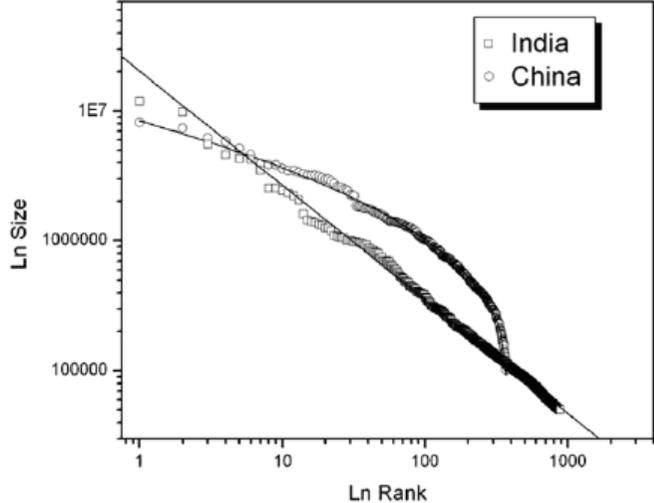 | B, A1 |
|---|---|---|---|---|
| 29. Cities | Rozenfeld, Rybski, Gabaix and Makse, 2010 | Partial Power law<br>Likely Lognormal | Cumulative<br>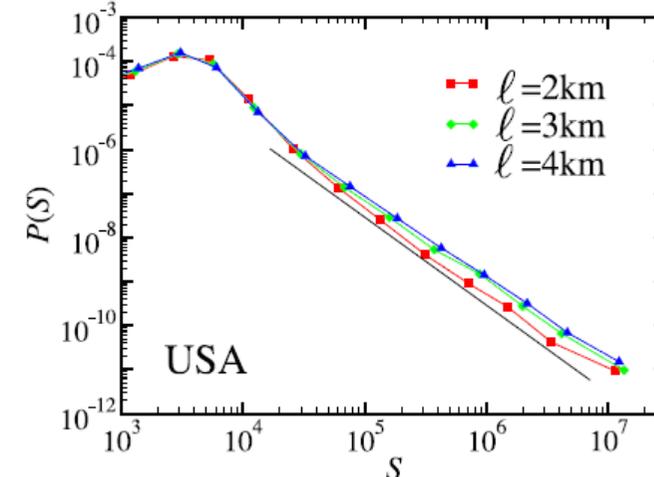 | A1 |

| 29. Cities | Berry and Okulicz-Kozaryn, 2011 | Power law | Rank-size 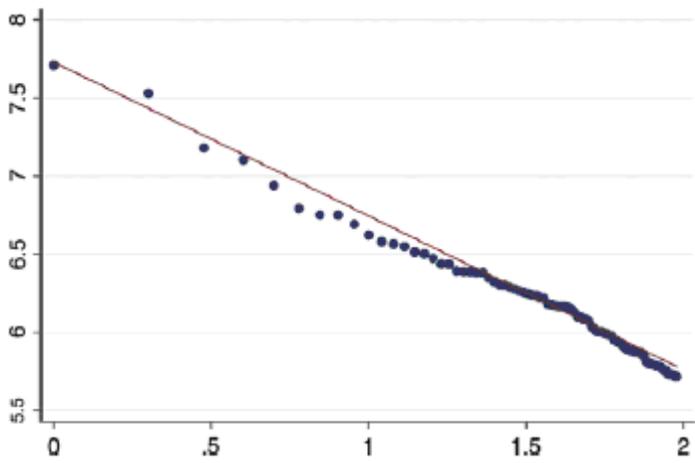 | B |
| --- | --- | --- | --- | --- |
| 29. Cities | Giesen and Suedekum, 2012 | Double Pareto lognormal (DPLN) | Density function 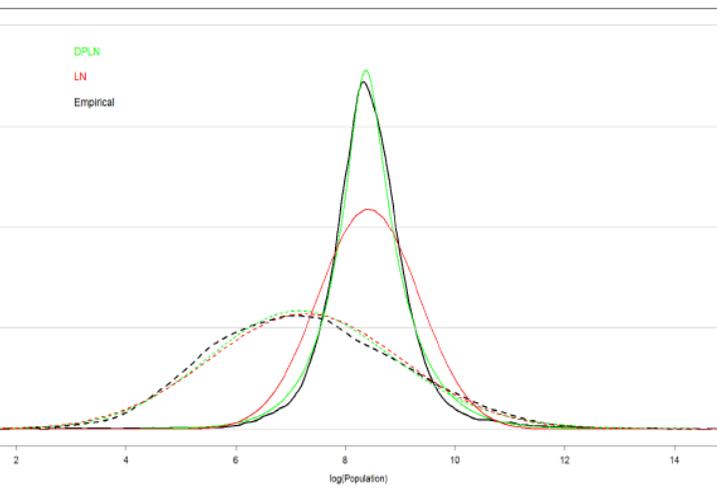 | A1 |
| 29. Cities | Ioannides and Skouras, 2012 | Most cities-lognormal, most of the population – power law | Rank –size 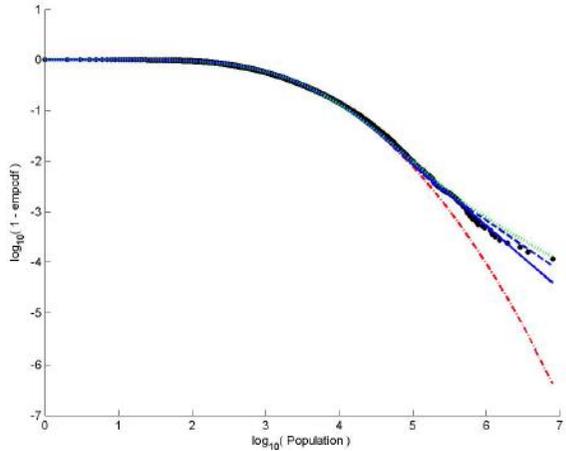 | C1 |
| **Linguistics** | | | | |
| **30. Frequency distribution of words** | Zanette and Montemurro, 2002 | Power law | Rank-size 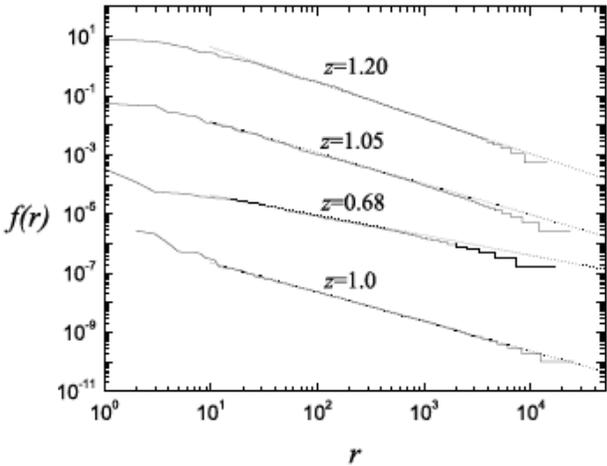 | B |

| 30. Frequency distribution of words | Montemurro, 2001 | Partial power law | Rank-size 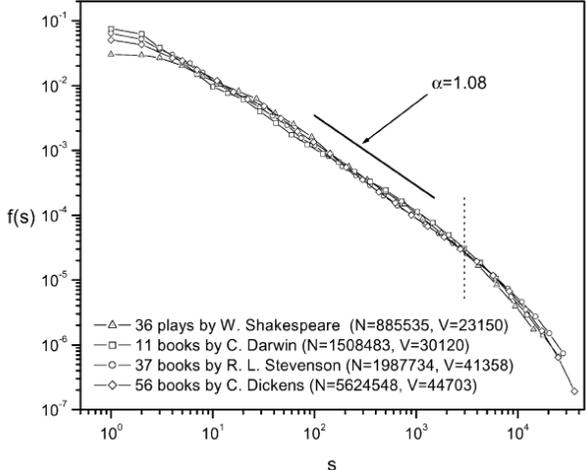 | C2 |
|---|---|---|---|---|
| 30. Frequency distribution of words | Piantodosi, 2014 | Partial Power law | Rank-size 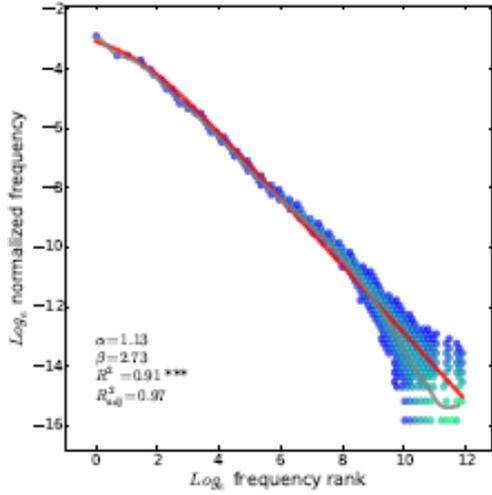 | C1 |

| 30. Frequency distribution of words | Dahui et al, 2005 | $p(k) = A - D ln(k)$ | Rank-size 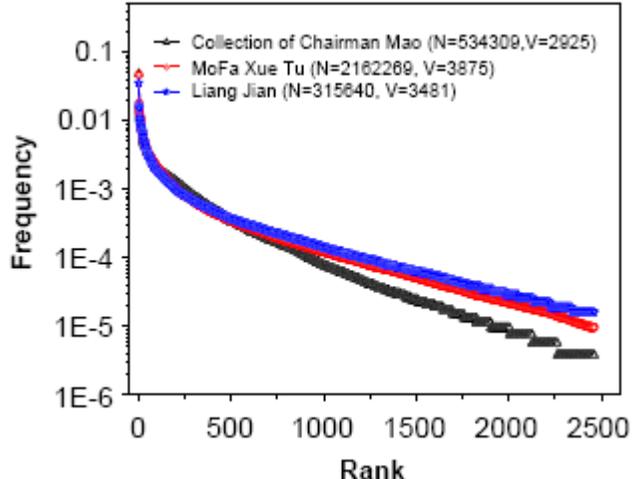 | D |
| --- | --- | --- | --- | --- |
| **31. Word length** | Sigurd et al, 2004 | Partial power law, gamma distribution | Density function 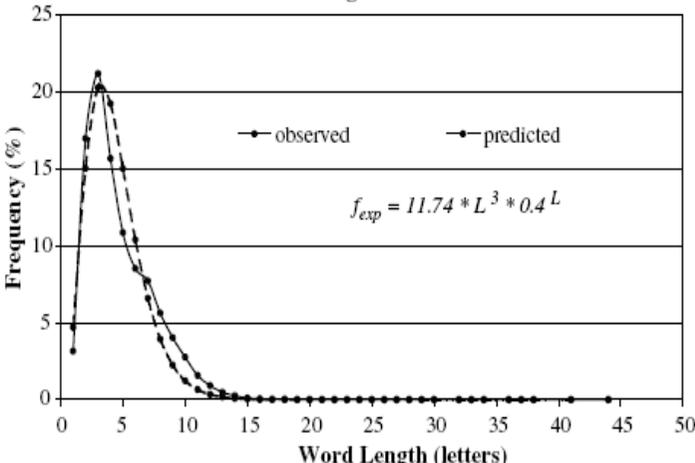 | A1 |

| 32. Language speakers | Stauffer et al 2006 | Lognormal | Histogram | A1 |
|---|---|---|---|---|
| **Economic activity** | | | | |
| 33. Firm sizes | Axtell, 2001 | Power law | Cumulative | B |

| 34. Strikes | Biggs, 2005 | Partial Power law | Cumulative 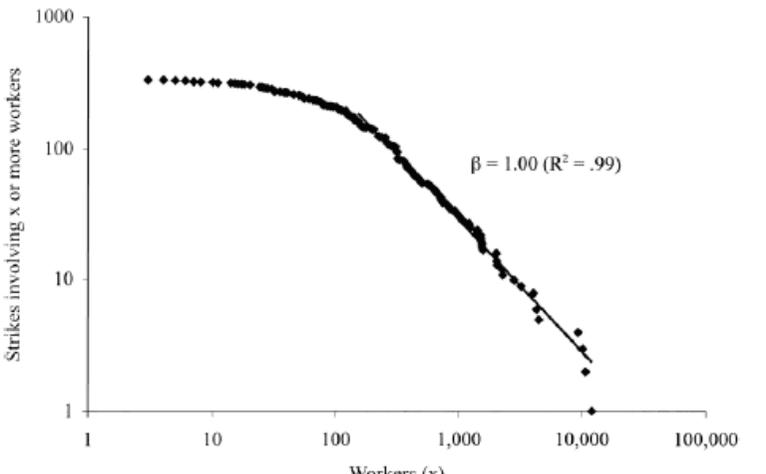 | A1 |
|---|---|---|---|---|
| 35. Income of companies | Okuyama, Takayasu, Takayasu, 1999 | Partial Power law | Cumulative 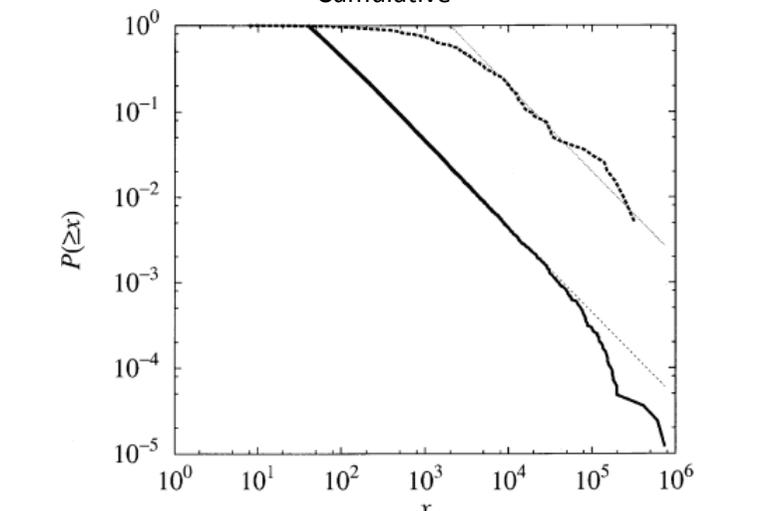 | B |

| 36. Incomes | Reed, 2003 | Double Pareto - 2 power laws | Histogram 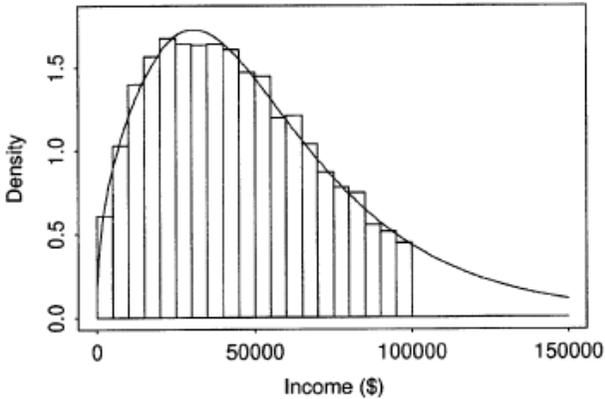 | A1 |
|---|---|---|---|---|
| 36. Personal income in Australia and US | Clementi, Di Matteo, Gallegati, Kaniadakis, 2008 | k-generalized distribution $$p(x) = \frac{\alpha \beta x^{\alpha-1} \exp_\kappa(-\beta x^\alpha)}{\sqrt{1+\kappa^2 \beta^2 x^{2\alpha}}}.$$ | Histogram 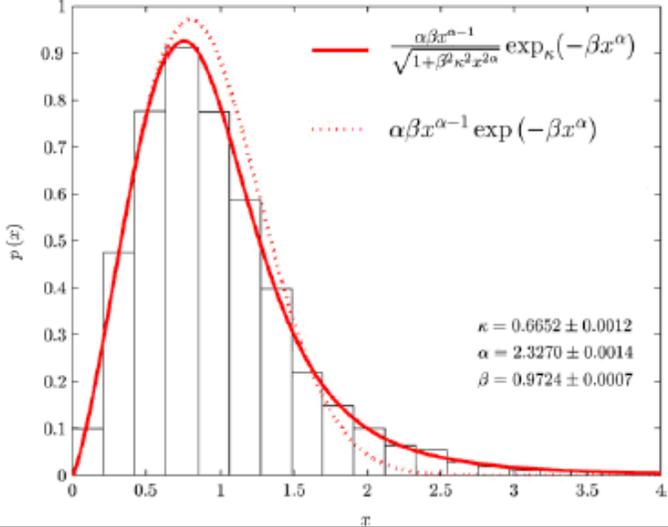 | A1 |

| 36. Individual income in US | Dragulescu and Yakovlenko, 2001 | Exponential distribution | Histogram 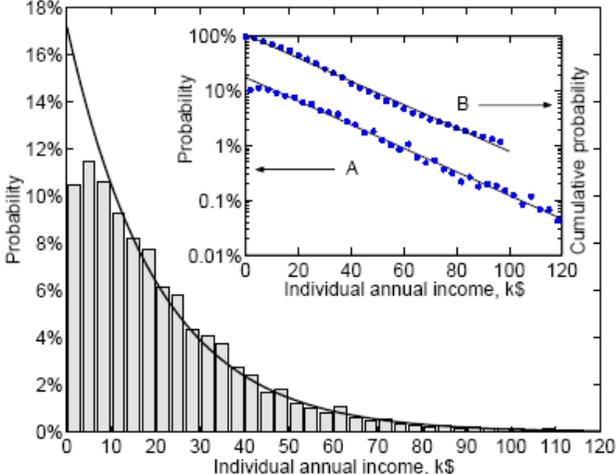 | A2 |
| --- | --- | --- | --- | --- |
| 36. Personal income | Lukasiewicz and Orlowski, 2004 | Dagum | Histogram 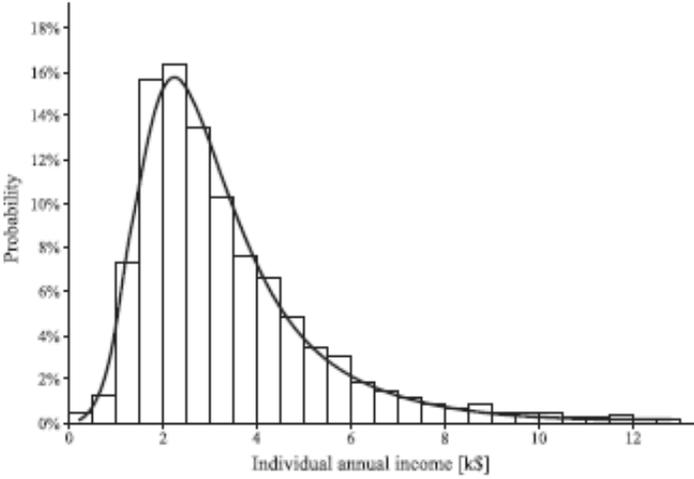 | A1 |

| 36. Incomes | Montroll and Schlesinger, 1983 | Power law – 2 different power laws | Cumulative 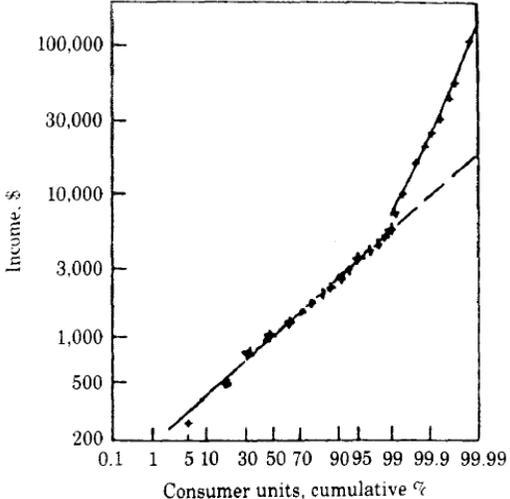 | E |
| --- | --- | --- | --- | --- |
| **37. Wealth (Forbes list of the richest people in the US)** | Levy and Solomon, 1997 | Power law | Rank-size 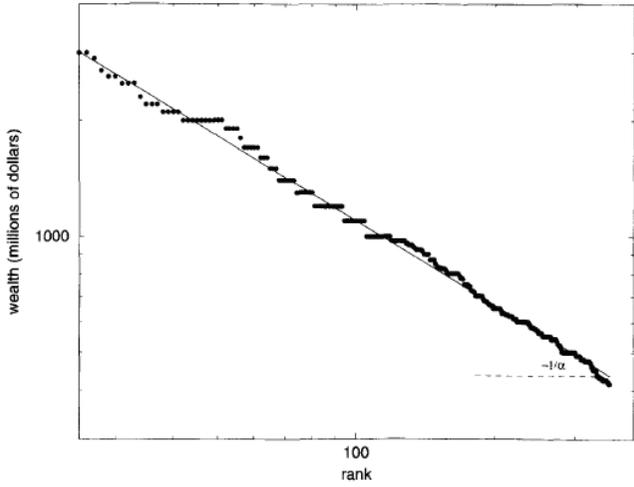 | B |

| 37. Wealth | Sinha, 2006 | Partial power law | Rank-size | B |
|---|---|---|---|---|
| **38. Income tax** | Fujiwara et al, 2003 | Power law (only for 1992) | Cumulative | B (1992) C1(1991) |

| 39. Size of shopping centers | Marinov, Benguigui and Czamanski, 2009 | Exponential distribution | Cumulative 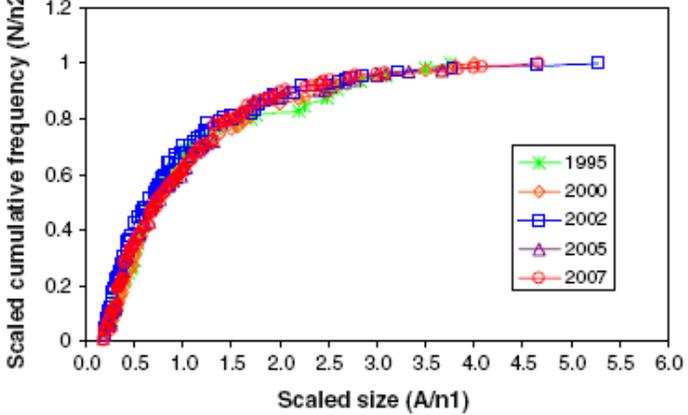 | A2 |
|---|---|---|---|---|
| 40. Connectivity of the electric power grid | Watts, 1999 | Exponential function | Cumulative 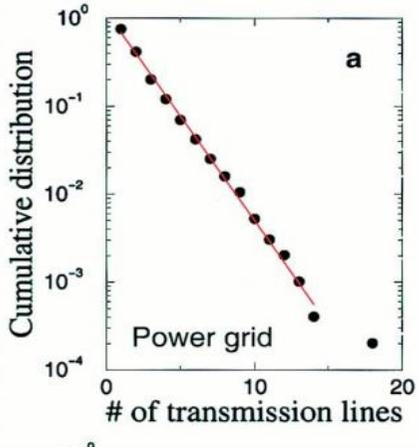 | A2 |